\documentclass[12 pt, amssymb,prb,showpacs]{revtex4}
\usepackage{epsfig}
\usepackage{dcolumn}
\usepackage{amsmath}
\begin{document}
\title{\bf Hysteretic behavior in weakly coupled double-dot transport in the spin blockade regime.}
\author{J. I\~narrea}
\affiliation{Escuela Polit\'ecnica Superior, Universidad Carlos III,
Leganes, Madrid, 28911, Spain\\ Instituto de Ciencia de Materiales,
CSIC, Cantoblanco, Madrid, 28049, Spain}
\author{C. Lopez-Monis}
\affiliation{Instituto de Ciencia de Materiales, CSIC, Cantoblanco,
Madrid, 28049, Spain}

\author{A.H. MacDonald} \affiliation {Department of Physics, University of Texas at Austin.Austin, Texas
78712}

\author{G. Platero} \affiliation{Instituto de Ciencia de Materiales,
CSIC, Cantoblanco, Madrid, 28049 Spain}

\begin{abstract}
Double quantum dot systems in the spin blockade regime exhibit
leakage currents that have been attributed to the Hyperfine
interaction. We model weakly coupled double-dot transport using a
rate equation approach which accounts for Hyperfine flip-flop
transitions. The rate equations allow us to obtain self-consistently
the time evolution for electronic charge occupations and for the
nuclei polarizations in each dot. We analyze the current in the spin
blockade region as a function of magnetic field and observe
hysteretic behavior for fields corresponding to the crossing between
triplet and singlet states.
\end{abstract}

\maketitle
\newpage
The Pauli exclusion principle can play an important role\cite
{ono,johnson,koop,Petta,onoprl} in current rectification in
molecular and semiconductor nanostructure transport. Spin blockade
(SB) is one important example which occurs in double quantum dots
(DQDs) over certain ranges of gate voltage, external field, and bias
voltage. The interplay between Coulomb and SB can be used to block
current in one direction of bias while allowing it to flow in the
opposite one. Because of this property DQDs can behave as externally
controllable spin-Coulomb rectifiers that have potential
applications in spintronics\cite{ono}. However, spin relaxation
processes, induced by spin-orbit or Hyperfine (HF) interactions,
produce a leakage current which partially removes SB. Recently,
striking features have been observed in tunneling spectroscopy
experiments in both lateral and vertical DQD's\cite{ono}, where spin
flip is attributed to HF interaction. In particular, hysteretic
current behavior as a function of an external magnetic field and
current instabilities, including time dependent current
oscillations, have been observed. These features were observed in
the SB regime, i.e., at bias voltages where the current is
drastically reduced.

In this letter, we model recent experimental studies of transport
through two weakly coupled vertical QD's \cite{tarucha,pfund}. These
experiments analyze transport through weakly coupled QD's in the
spin blockade regime under an external magnetic field ($B$). In the
experimental current versus $B$ curves, two striking features are
observed: an step in the current and hysteretic behavior. The
current step position depends on the source-drain voltage
($V_{DC}$). We propose a model based on rate equations which
includes a microscopic approach of the Hyperfine
interaction\cite{ina}. The time evolution for electronic charge
occupations and nuclei polarizations in each dot is obtained by
means of rate equations which are self-consistently solved. In our
model we consider up to two extra electrons in the system. Double
occupation is allowed only in the right QD. As well we assume that
HF interaction is different for the left and right QD. It would be
the case for instance, when the number of nuclei within each dot is
different. The DQD is biased by a $V_{DC}$ which brings into
resonance the inter-dot two electrons state, with opposite spin for
each electron, with the right intra-dot singlet state. In this
situation the current flows till the electrons in the left and right
QD have the same spin polarization. The current drops until spin
flip occurs. Then, just a small leakage current flows.

We consider a Hamiltonian: $H=
H_{L}+H_{R}+H_{T}^{LR}+H_{leads}+H_{T}^{l,D}$ where $ H_{L} (H_{R})$
is the Hamiltonian for the isolated left (right) QD and is modelled
as one-level (two-level) Anderson impurity.
$H_{T}^{LR}$($H_{T}^{l,D}$) describes tunneling between QD's
(superscript "l" describes leads and  "D" QD's)\cite{cota} and
$H_{leads}$ is the leads Hamiltonian. The present model uses rate
equations for electron states occupation probabilities. We neglect
coherences (reversible dynamics), which is a reasonable assumption
as the dots are weakly coupled. The model includes rate equations
also for the mean polarization of the nuclei in each QD\cite{ina}.
Spin flip is included by means of a microscopic model for HF
interaction and finally rate equations are self-consistently solved.
Our basis consists of twenty states, but those which mostly
participate in the electron dynamics at the SB region are:
$|\uparrow,\uparrow\rangle$,
$|\downarrow,\downarrow\rangle$, $|\uparrow,\downarrow\rangle$,
$|\downarrow,\uparrow\rangle$ and $|0,\uparrow\downarrow\rangle$.

Rate equations for state occupation probabilities $\rho_{s}$ are:
\begin{equation}
 \dot{\rho}(t)_{s} =\sum_{m\neq s} W_{sm}\rho_{m} -
\sum_{k\neq s} W_{ks}\rho_{s}
\end{equation}
where $W_{i,j}$ is the transition rate \cite{ina} from state $j$ to
state $i$ . The HF Hamiltonian is:
\begin{equation}
 \hat{H}= \hat{H}_{z}+\hat{H}_{sf}
\end{equation}
 where
\begin{equation}
\hat{H}_{z}= [ A \langle I_{z} \rangle ] S_z
\end{equation}
On the other hand,
\begin{equation}
\hat{H}_{sf}= (A/2N) \sum_{i} \left[
S_{+}I_{-}^{i}+S_{-}I_{+}^{i}\right]
\end{equation}
is the flip-flop interaction responsible for mutual electronic and
nuclear spin flip. $A$ is the average HF coupling constant and the
nuclear spin $I$=1/2. Because of the mismatch between nuclear and
electronic Zeeman energies transitions must be accompanied at low
temperature by phonon emission. We approximate the electronic sf
transition rate as\cite{ina}:
\begin{equation}
\frac{1}{\tau_{sf}}\simeq\frac{2\pi}{\hbar} |< \hat{H}_{sf}>|^{2}
\frac{\gamma}{\Delta E^{2}+\gamma^{2}}
\end{equation}
where $\gamma$ is the electronic state life-time broadening which is
of the order of $\mu eV$, i.e., of the order of the phonon
scattering rate \cite{fujisawa}.
$\Delta E$ is the difference between the energy of a state with one
electron in each dot with aligned spins
($|\downarrow,\downarrow\rangle$/$|\uparrow,\uparrow\rangle$)and the
energy of a state with one electron in each dot with opposite spin
orientation
($|\uparrow,\downarrow\rangle$/$|\downarrow,\uparrow\rangle$)  (see
Fig. 1). The latter is $mixed$ due to interdot tunneling with the
intradot singlet state in the right QD
($|0,\downarrow\uparrow\rangle$). The energy of the $mixed$ state is
calculated through a {\it two level system} approach and depends
mainly on the inner barrier coupling term ($t$) and the right and
left dots level detuning ($\Delta$). Detuning, $\Delta$, is defined
as the difference between the energy of a state with one electron in
each dot with aligned spins and the energy of a state with two
electrons with opposite spin orientation in the right QD,
($|0,\downarrow\uparrow\rangle$):
\begin{equation}
\Delta=E_{(|\downarrow,\downarrow\rangle/|\uparrow,\uparrow\rangle)}-
E_{(|0,\downarrow\uparrow\rangle)}
\end{equation}
At $B\neq 0$, $\Delta E$ depends on $B$ and on the nuclei spin
polarization:
\begin{equation}
\Delta E=
E_{(|\downarrow,\downarrow\rangle/|\uparrow,\uparrow\rangle)}-
E_{(|\uparrow,\downarrow\rangle/|\downarrow,\uparrow\rangle)}
\propto g_{e}\mu_{B} B +\frac{A}{2}P
\end{equation}
Here $P$ characterizes the nuclear spin configuration and is defined
by $P= \frac{(N_{1/2}-N_{-1/2})}{N}$, where $N_{1/2}$ is the number
of nuclei with $I_{Z}=1/2$ and $N_{-1/2}$ is the number of nuclei
with $I_{Z}=-1/2$. We consider a finite nuclear spin relaxation
 time due to nuclei spin scattering
($\approx $ ms \cite{fujisawa,mer}). The system of time evolution
equations for the electronic states occupations $ \rho_i $ and
nuclei polarization of the left and right dot is self-consistently
solved. From that we calculate the total current through the system
which is the physical observable of interest (see ref. \cite{ina}).

In Fig. 1 the energy levels diagram as a function of $\Delta$ is
shown. At finite $B$ the interdot triplet state splits (see bottom
panel of Fig. 1). At large detuning, increasing $B$,
$|\downarrow,\downarrow\rangle$ is close to the
$|\uparrow,\downarrow\rangle$  but electron-nuclei spin scattering
is not efficient because it would imply phonon absorption which has
very low probability at low temperature. When
$|\downarrow,\downarrow\rangle$ crosses the
$|\uparrow,\downarrow\rangle$ ( $|\downarrow,\uparrow\rangle$)
state, $\Delta E=0$  and according to eq. (5) spin flip has the
largest probability to occur. Now electrons and nuclei spin flip
processes takes place through HF interaction with phonon emission.
They are favorable at temperature close to 0. This is the physical
origin for the current step experimentally
observed\cite{tarucha,pfund}. The calculated current versus magnetic
field is presented in Fig. 2 where the current step at different
source-drain voltages is shown. As the level crossing occurs the
current flows and a finite nuclei polarization is induced (see Figs.
2 and 3). It produces an additional Zeeman term which re-normalizes
the energy levels. Increasing further $B$, the current remains
constant due to the interplay between the flip flop processes and
spin scattering between nuclei which acts removing the nuclei
induced spin polarization. Sweeping $B$ backwards, the current
remains finite up to the crossing of the levels which now takes
place at lower $B$ than in the sweeping forward case due to the
feedback between charge occupation and nuclei spin polarization (see
Fig. 4). Therefore the current presents a clear hysteretic behavior.
We observe as well how the bistability region depends on $V_{DC}$
(as experimentally observed).

In conclusion we have analyzed charge transport through weakly
coupled DQDs in the SB regime including HF interactions and
considering phenomenologically phonon emission as dissipative
mechanism. SB is removed at triplet-singlet levels crossing, once
flip-flop mechanism is assisted by phonon emission. The interplay
between HF interaction, nuclei dipole interaction and electronic
charge occupation produces bistability in the current as a function
of the external magnetic field.

This work has been supported by the MCYT (Spain) under
grant MAT2005-06444 (JI and GP), by the Ram\'on y Cajal program
(J.I.) by the EU Programme: HPRN-CT-2000-00144, by
the Welch Foundation (AHM) and by the DOE (AHM) under grant
DE-FG03-02ER45958.
\newpage


\newpage
\clearpage

Figure 1. Caption: Bottom panel: Energy levels diagram as a function
of the detuning at finite $B$. At large detuning
$|\downarrow,\downarrow\rangle$ is close to the
$|\uparrow,\downarrow\rangle$  but electron-nuclei spin scattering
is not efficient because it would imply phonon absorption which has
very low probability at low temperature. Increasing $B$, both states
cross. As $|\downarrow,\downarrow\rangle$ crosses the state
$|\uparrow,\downarrow\rangle$, spin flip flop between electron and
nuclei occurs and the current begins to flow. Top panel: same as
bottom panel with $B=0.$
\newline

Figure 2. Caption: Current versus magnetic field $B$. Increasing
$B$, as the level crossing occurs the current flows and a finite
nuclei and electron spin polarization is induced. It produces an
additional Zeeman term which re-normalizes the energy levels and the
current begins to flow. Increasing further $B$, the current remains
constant due to the interplay between the flip flop processes and
spin scattering between nuclei which acts removing the nuclei
induced spin polarization. Sweeping $B$ backwards, the current
remains finite up to the crossing of the levels which now takes
place at lower $B$ due to the induced electron and nuclei spin
polarization
\newline

Figure 3. Caption: Nuclear polarization $P$ versus sweeping up and
down $B$. It can be observed the hysteretic behavior in $P$ giving
rise to bistability regions.
\newline

Figure 4. Caption: $\Delta E$ versus sweeping up and down $B$ at
source-drain voltage of 6.6 meV.
\newpage

\begin{figure}
\centering\epsfxsize=4.5in \epsfysize=5.5in \epsffile{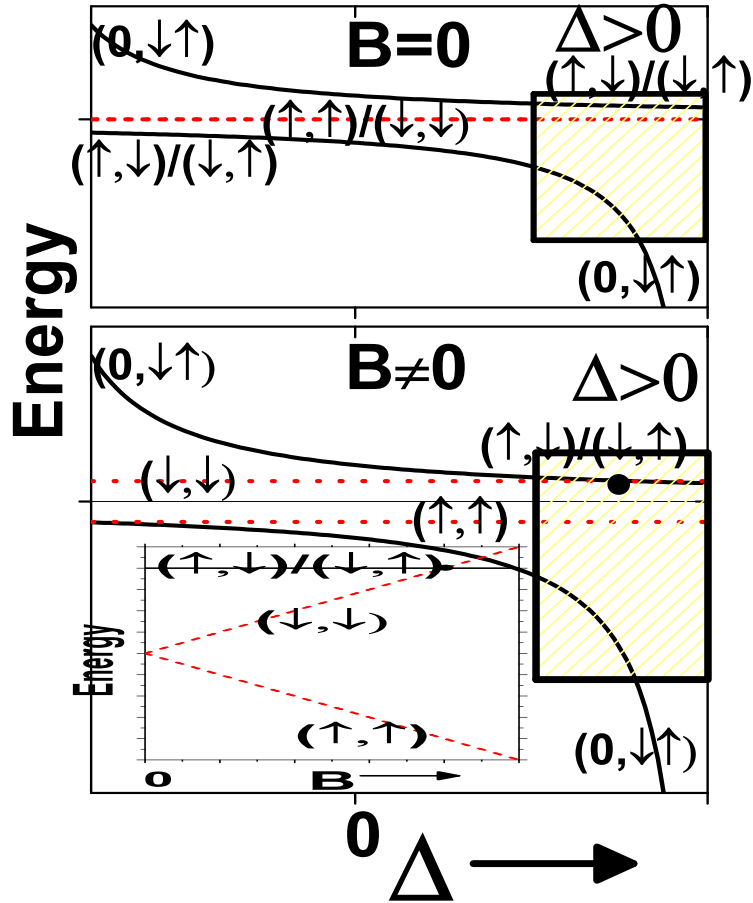}
\caption{ Bottom panel: Energy levels diagram as a function of the
detuning at finite $B$. At large detuning
$|\downarrow,\downarrow\rangle$ is close to the
$|\uparrow,\downarrow\rangle$  but electron-nuclei spin scattering
is not efficient because it would imply phonon absorption which has
very low probability at low temperature. Increasing $B$, both states
cross. As $|\downarrow,\downarrow\rangle$ crosses the state
$|\uparrow,\downarrow\rangle$, spin flip flop between electron and
nuclei occurs and the current begins to flow. Top panel: same as
bottom panel with $B=0.$}
\end{figure}

\newpage
\clearpage

\begin{figure}
\centering\epsfxsize=4.5in \epsfysize=5.0in \epsffile{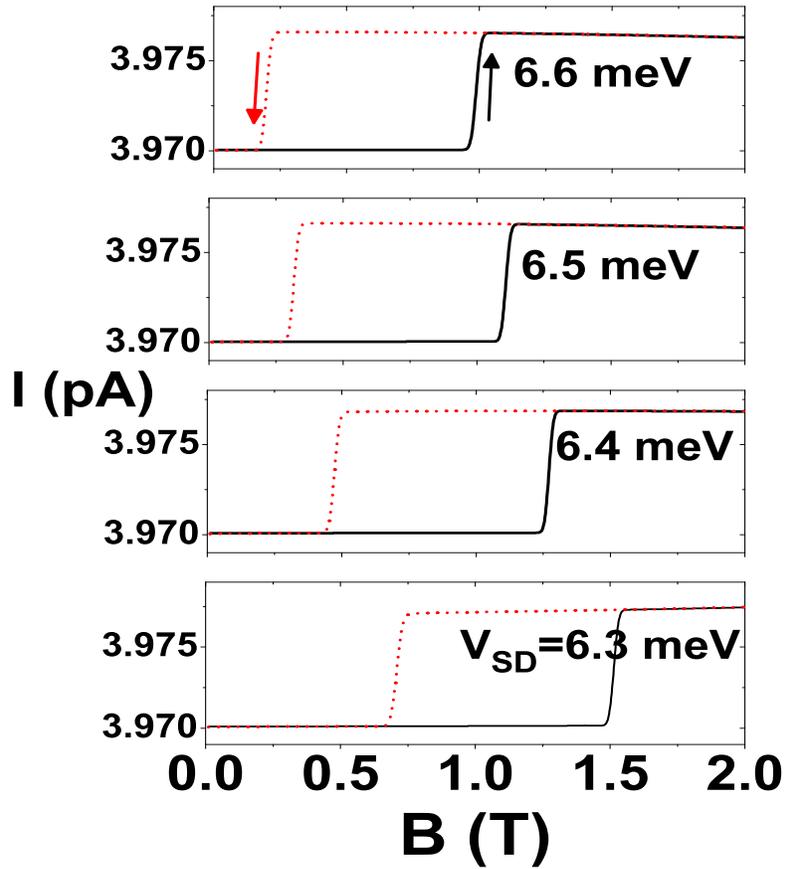}
\caption{ Current versus magnetic field $B$. Increasing $B$, as the
level crossing occurs the current flows and a finite nuclei and
electron spin polarization is induced. It produces an additional
Zeeman term which re-normalizes the energy levels and the current
begins to flow. Increasing further $B$, the current remains constant
due to the interplay between the flip flop processes and spin
scattering between nuclei which acts removing the nuclei induced
spin polarization. Sweeping $B$ backwards, the current remains
finite up to the crossing of the levels which now takes place at
lower $B$ due to the induced electron and nuclei spin polarization }
\end{figure}

\newpage
\clearpage

\begin{figure}
\centering\epsfxsize=4.5in \epsfysize=5.0in \epsffile{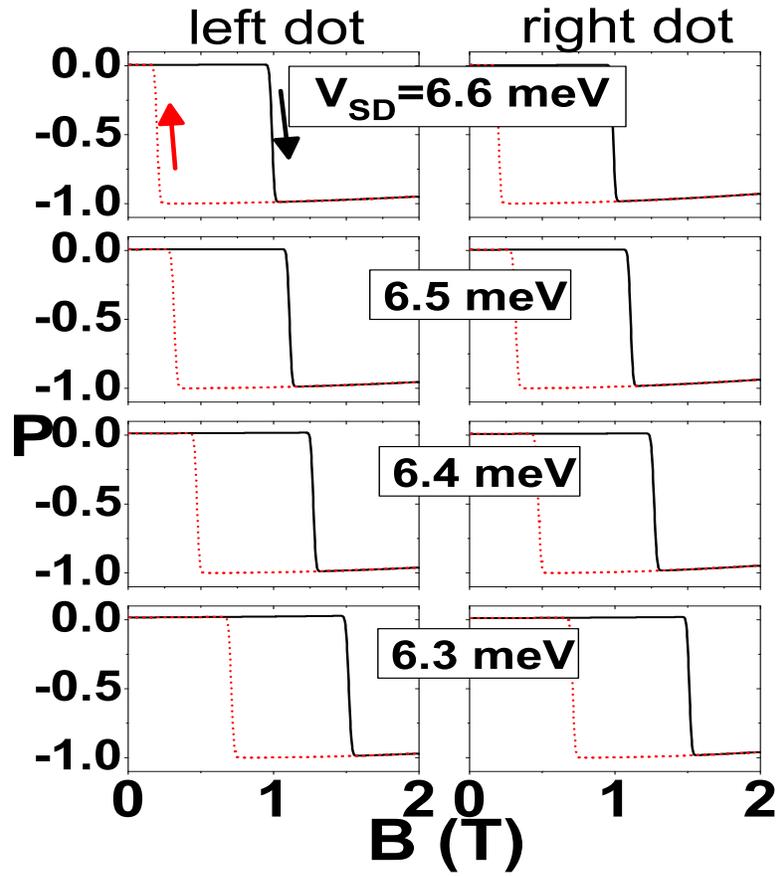}
\caption{ Nuclear polarization $P$ versus sweeping up and down $B$.
It can be observed the hysteretic behavior in $P$ giving rise to
bistability regions.}
\end{figure}

\newpage
\clearpage

\begin{figure}
\centering\epsfxsize=4.5in \epsfysize=4.5in \epsffile{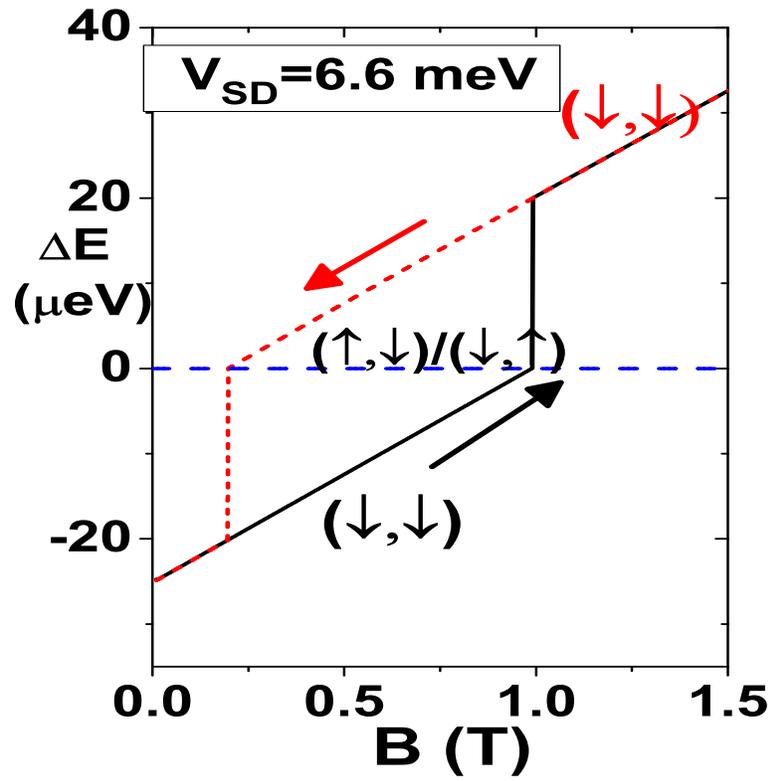}
\caption{ $\Delta E$ versus sweeping up and down $B$ at source-drain
voltage of 6.6 meV.}
\end{figure}

\end{document}